\documentclass{aa}
\usepackage{natbib}
\usepackage{graphicx}

\def\spose#1{\hbox to 0pt{#1\hss}}
\def\lta{\mathrel{\spose{\lower 3pt\hbox{$\mathchar"218$}}
        \raise 2.0pt\hbox{$\mathchar"13C$}}}
\def\gta{\mathrel{\spose{\lower 3pt\hbox{$\mathchar"218$}}
        \raise 2.0pt\hbox{$\mathchar"13E$}}}

\begin{document}

\title{QPOs in Cataclysmic Variables and in X-ray Binaries}

\author{W{\l}odek Klu\'zniak\inst{1,2,3}
\and Jean-Pierre Lasota\inst{1,4} \and Marek A.
Abramowicz\inst{1,5}
\thanks{Address: Theoretical Physics, Chalmers
University, S-412-96 G\"oteborg, Sweden}
\and Brian Warner\inst{6} }
\offprints{J.-P. Lasota}
\institute{ Nordita, Blegdamsvej 17,
DK-2100, Copenhagen, Denmark \and Institute of Astronomy, Univ. of
Zielona G\'ora, ul. Lubuska 2, PL-65-265
Zielona G\'ora, Poland
\and Copernicus Astronomical Center, ul.
Bartycka 18, PL-00-716
Warszawa, Poland
\and Institut d'Astrophysique de Paris,
UMR 7095 CNRS, Univ. P. \&
M. Curie, 98bis Bd Arago, 75014 Paris, France
\and Department of Astrophysics, G\"oteborg
University,
G\"oteborg, Sweden
\and
Department of Astronomy, University of Cape Town, Rondebosch 7700, South Africa
}
\date{Received; accepted}
\titlerunning{QPOs in CVs and LMXBs}

\abstract{Recent observations, reported by Warner and Woudt, of
Dwarf Nova Oscillations (DNOs) exhibiting frequency drift, period
doubling, and 1:2:3 harmonic structure, can be understood as disc
oscillations that are excited by perturbations at the spin frequency
of the white dwarf  or of its equatorial layers. Similar
quasi-periodic disc oscillations in black hole low-mass X-ray binary
(LMXB) transients in a 2:3 frequency ratio show no evidence of
frequency drift and correspond to two separate modes of disc
oscillation excited by an internal resonance. Just as no effects of
general relativity play a role in white dwarf DNOs, no stellar
surface or magnetic field effects need be invoked to explain the
black hole QPOs.

\keywords{ accretion, accretion discs -- novae, cataclysmic
variables -- X-ray: binaries} }
\maketitle

\section{QPO similarities and differences between white dwarfs and
black holes}

The highest frequencies of nearly periodic modulations of  X-rays
(HF QPOs) in  black holes and neutron stars  \citep[][and references
therein]{vdk} attract a great deal of attention because it is
thought that their frequencies reflect dynamic phenomena in the
motion of matter  in strong-field gravity. Disc oscillations in
Einstein's gravity are a favored explanation \citep{kato01,bob99},
and a 2:3 ratio of frequencies has been pointed out by \citet{ak32}
and suggested to be a manifestation of non-linear internal resonance
of accretion discs.

Recently, \citet[][and references therein]{ww2} have discovered that
in dwarf nova outbursts  in VW Hyi, the DNO oscillations (sub-Hz
dwarf nova oscillations reminiscent of HF QPOs in neutron stars and
black holes) exhibit period doubling and tripling. At times two or
three frequencies in a 1:2:3 ratio are present at the same time in
the light curve. At first sight this seems analogous to the
frequencies reported in black holes \citep[e.g., the frequencies in
a 1:2:3 ratio in the source XTE J1550-564;][]{rem02}, and led to
attempts at finding a common model for DNOs in accreting white
dwarfs and QPOs in candidate black holes \citep{ww2}. However, while
the two phenomena are similar, they are not identical.

The differences are two-fold. In black holes binaries the twin HF
QPOs have fixed frequencies, within errors of several percent
\citep{mcr}, while the DNO frequencies vary with time (as the
luminosity decreases) by a factor exceeding two. Further, in white
dwarf DNOs, on occasion, two or more variable frequencies are
present simultaneously, always in a 1:2:3 ratio, i.e., more than one
harmonic is present. In black holes, usually only one mode of
oscillation is manifest at a given time. Although sometimes both HF
QPO frequencies are present at the same time in some black holes
\citep{str01,rem02}, careful analysis reveals that the two
frequencies are {\it not} harmonics of one non-sinusoidal
oscillation \citep{mcr}.

We suggest that the DNOs can be interpreted in terms of resonant
disk oscillations invoked to explain the HF QPOs in black hole
binaries \citep{ak32,ka03}, but with another excitation mechanism
that is related to the difference between black holes and white
dwarfs. White dwarfs have a surface which can, e.g., support a
magnetic field, and is capable of disturbing the disc strictly
periodically at a well-defined rotation rate. None of these effects
is present in black holes.

\section{QPO correlations and frequency scalings}

We digress to note that the mechanism of QPO formation has
previously been suggested to be common to sources as diverse as
black holes and white dwarfs based on a linear relationship between
two frequencies in LMXBs \citep{WK99, PBK99, BPK02}, later extended
to  cataclysmic variables \citep{mauche,ww1}. Because orbital motion
around white dwarfs is accurately described by Newtonian gravity
this seemed to rule out the few models, such as the so called
relativistic precession model, in which {\it all} QPO frequencies
can be related to general relativistic frequencies
\citep{mauche,ww1}. However, HF QPOs in black holes do not appear
together with a low-frequency QPO that would allow them to be placed
on the  correlation (Belloni 2005, private communication). In
another view, the general scaling of frequencies with radius and
mass, apparent in QPO sources, and anticipated in \citet{kmw90},
suggests that QPOs and DNOs are an accretion disc phenomenon
\citep{kal04}.

\section{A disc oscillation model for white dwarf DNO{\small s}.}

It has been suggested that the twin HF QPOs in black holes
correspond to two different modes of disc motion (e.g., radial
oscillations and essentially vertical oscillations of the disc)
which are in a 2:3 frequency ratio because they are excited by an
internal resonance in the accretion disc \citep{ka03}. The first
mode can modulate the emissivity of the disc, but the second mode is
less likely to do so. However, in a black hole, axisymmetric
vertical motion of the disc can modulate the X-ray luminosity
through gravitational lensing at the source, because the light
trajectories are bent by differing amounts for different positions
of the disc \citep{bur04}. One of the two modes of oscillation
occurs at the radial epicyclic frequency---or rather, its value at a
certain position close to that of the pressure maximum of the
accretion disc: \cite{zanotti03,rubio05}---and the other at the
vertical epicyclic frequency, in the same sense \citep{lee04}. We
will call the epicyclic frequencies at this position in the disk
``central.''The two mode frequencies are different in strong-field
Einstein's gravity, but they are equal in Newtonian $1/r$ potential.
The presence of two frequencies in white dwarfs cannot be explained
by excitation of two distinct  disc oscillation modes that are
degenerate in frequency---harmonic overtones are a more likely cause
in these dwarf novae\footnote{The radial overtones are harmonic as
in a flute mode \citep{rezz}.}.

\begin{figure}
\resizebox{\columnwidth}{!}
{\includegraphics[width=84mm]{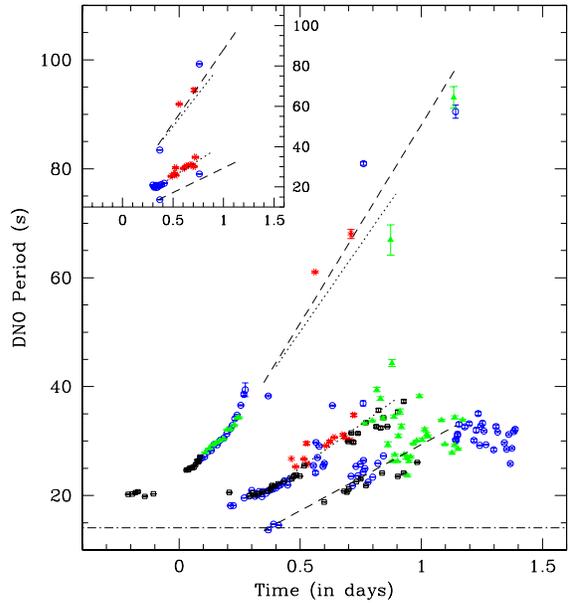}}
\caption{The evolution
of DNO periods at the end of normal and super outbursts in the dwarf
nova VW Hyi. The different symbols indicate the various different
kind of outbursts (short: asterisk, normal: open circles, long: open
squares, and super outbursts: filled triangles). The dotted and
dashed lines show the result of a least-squares fit to the first and
second harmonic, respectively, and after a rescaling in period by a
factor of two or three, respectively, they are replotted in order
better to show the evolution of the DNO period. The zero of outburst
phase is defined in \citet{ww02}. The inset highlights two observing
runs in which the fundamental, first and second harmonic of the DNO
period were present simultaneously. The horizontal dotted-dashed
line illustrates the minimum DNO period (14.1 s) observed at maximum
brightness. (From \citet{ww2}).} \label{warnerf4}
\end{figure}

An analysis of the physical and statistical properties of the twin
HF QPOs in neutron star systems has led to the suggestion that
identical modes are resonantly excited in the neutron star and black
hole systems \citep{ka00,ka01,bulik}. It has also been pointed out
that in neutron stars an additional source of disc excitation is
present at its center, a non-axisymmetric rotating magnetosphere
\citep{kal04,lee04,kato05}, and that direct evidence of a resonance
excited in this way is present in the transient accreting X-ray
pulsar SAX J1808.4-3658 \citep{wij,kluetal}. Here, we note that the
same source of excitation should be present in white dwarf systems,
if a magnetospheric structure is present. Indeed, the observed
period-luminosity relationship for DNOs prompted \citet[][see also
Warner 1995]{BP} to suggest that magnetically channelled accretion
was responsible, whereas \citet{king} attributed DNOs to the
presence of transient magnetic fields generated by turbulent dynamo
in the white dwarf's outer layers.

We suggest two alternative explanations of the frequency evolution
in Dwarf Nova Oscillations, based on extensive post-outburst
observations of VW Hyi \citep{ww2}, see Fig.~1. The primary of the
cataclysmic variable VW Hyi is a white dwarf of mass between $0.6$
and $0.8 M_\odot$, which would correspond to a radius between 8 and
6 $\times10^3\,$km \citep{schvgt,sionetal97}. The rotation period is
not known, but spectral fits in  quiescence suggest two components
to the observed surface velocity ($v\sin i$), about 400 km/s and
about 4000 km/s \citep{szkody}, this could correspond to a spin  of
$\sim10^2\,$s or rotational period of an equatorial accretion belt
of $\sim10\,$s. At maximum of outbursts DNOs are rarely seen, but
when they are they are at 14.1 s \citep{ww2}.

Warner and Woudt (2005a) point out that the frequency of DNOs
decreases with the mass accretion rate and interpret this as the
pushing out of the inner edge of the accretion disc by a magnetic
field structure whose pressure increasingly overcomes the ram
pressure of the accretion flow. In another interpretation of the DNO
fundamental frequency, \citet{ww3} suggest that the  frequency,
linearly decreasing in time, can be understood as the period of
rotation of an equatorial accretion belt magnetically coupled to the
accretion disc, and possessing a non-axisymmetric structure (for the
model, originally proposed by \citet{BP}, see \citet{w95}). We
suggest that the evolution of the DNO frequency (plotted in Fig.~1)
can be understood as a resonant response of the accretion disk to
periodic perturbations by a magnetic field structure. In the first
interpretation the perturbation period is fixed, but the disk
eigenfrequency decreases. In the second interpretation the
perturbation period decreases in time, while the eigenfrequency of
the disk does not change.

An inspection of the data of Warner and Woudt (see Fig.
\ref{warnerf4}) reveals that the frequency doubling does not occur
at an arbitrary moment, but at that specific instant when the DNO
frequency is about {\it one half} of its initial frequency. This can
be understood if the observed frequency is that of an oscillation
excited in the accretion disc by a perturbation related to the spin
of the white dwarf, or of its equatorial layers. During the decay
from outburst maximum either a permanent dipole structure of the
white dwarf \citep{BP,jpl}, or a transient structure generated in a
mechanism suggested by \citet{king}, affects the inner disc. There
is an increasing mismatch of the perturbing frequency and the
eigenfrequency of the disk, eventually leading to period halving, as
described below.

Under the first interpretation, as the mass accretion rate drops
(with the time elapsed from outburst maximum), the magnetic pressure
pushes out the inner radius of the accretion disc, as in
\citet{ww2}, and the position of the pressure maximum in the disc
moves out accordingly. The frequency of the fundamental mode
decreases with the decreasing Keplerian frequency in the relevant
parts of the disc,\footnote{Once again, recall that in a Newtonian
$1/r$ potential the epicyclic frequencies are equal to the orbital
frequency.} and so do the frequencies of the overtones. The crucial
point, is that the ratio of the overtones to the fundamental remains
fixed in the process.

All the while, the white dwarf disturbs the disc periodically at its
(the white dwarf) spin frequency. The coupling is sufficiently weak
that the direct forcing frequency is barely present in the disc,
instead, the disc responds to the perturbation at its own
eigen-frequency \citep{kluetal,lee04}, e.g., the ``central''
epicyclic frequency, which constantly decreases during the decay of
the dwarf-nova outburst. Eventually the eigenfrequency of the disc
drops to about one half of the perturbing frequency, and the
perturbation can now resonantly excite the first harmonic. Once
excited, this harmonic is preferentially maintained by the same
perturbations, even as the disc eigenfrequency decreases. In VW Hyi
at late times in the outburst, three frequencies are seen, in a
1:2:3 ratio (Fig.~1) The highest of these could be the second
overtone, or it could be the non-linear beating of the strongly
excited first overtone and the weakly present fundamental, at
$f_3=f_2+f_1$. The difficulty of this model lies in the period of
the white dwarf suggested by observations mentioned above
 \citep{szkody}, nearly an order of magnitude larger than the $\sim 20\,$s
period required for this model of period halving.

In the second interpretation, the perturbing structure rotates at a
variable rate along the equator, at rotation periods possibly as
short as $14s$, in agreement with the observations of
\citet{szkody}. The decreasing frequency would then reflect the
decreasing rate of angular momentum accretion onto the white dwarf,
but the relevant properties of the disc (such as its inner radius)
would not vary strongly in the relevant parts  of the outburst
(which is consistent with magnetic moment assumed in the VW Hyi
outburst model of \citet{shl04}). In this case, if one assumes
strong coupling of the boundary layer to the disc, we would expect
(and suggest) that the oscillatory motion of the disc is well
approximated by that of a forced non-linear oscillator, i.e., the
prime response would be at the forcing frequency of the boundary
layer rotation. As this forcing rotation decreases, eventually a
resonant response at a harmonic of the forcing frequency occurs,
when it becomes equal to about one-half (and later one-third, for
the second harmonic) of the accretion disc eigenfrequency.

We suggest that the observed DNO corresponds to the disc response.
In support of this interpretation, we recall that a non-linear
resonance occurs in a certain frequency range. In particular, as the
driver frequency decreases, the resonance turns on suddenly, and at
a frequency higher than the eigenfrequency of the oscillator. We
note that in the data of Fig.~1, the harmonics first appear at
shorter periods (larger frequencies) than the shortest period of the
fundamental, which was observed earlier in the outburst.

A similar mechanism cannot be invoked for black hole discs, not only
for lack of a suitable perturbing agent. If in black holes the two
HF QPOs in the 2:3 frequency ratio were also overtones, they could
be excited in a range of frequencies (just as the DNO overtones are
present over a range of frequencies, c.f., Fig.~1). Instead, the
black hole QPOs appear at fixed frequencies. This is consistent with
two oscillatory modes of an accretion disc whose frequency ratio
varies with the properties of the disc, and only for a certain state
of a disc has the value 2:3, which can be excited by an internal
resonance. For instance, the ratio of the vertical to radial
epicyclic frequencies varies smoothly between unity and infinity, as
the circle of maximum pressure in the disc moves in from very large
radii to the radius of the marginally stable orbit.

When a monotonic decay occurs as in the neutron-star X-ray transient
system Aql X-1 the (single) kHz QPOs are observed only during the
so-called transition state close to maximum and then only
short-timescale frequency-flux correlations are recorded
\citep{mb04,zhang}. In general, in neutron-star X-ray binaries no
QPOs are observed at the maximum and at the end of the outburst
\citep[e.g.,][]{cui}. One should keep in mind, however, that VW Hyi
is unique among CVs in having QPOs that are most prominent towards
the end of the outburst.

\section{Conclusions}

The frequency evolution of the DNO oscillations, and their harmonic
structure can be understood as the response of the disc oscillator
to an external perturbation applied at a constant frequency while
the disc eigenfrequency decreases with time, as its properties
change. Alternatively, the DNO could be  the forced response to a
perturbation of steadily increasing period of a disc with a fixed
eigenfrequency. The frequency drift is reminiscent of the kHz QPOs
in low-mass X-ray binary neutron stars, where the disc is also
perturbed at the stellar spin frequency. But unlike in persistent
LMXBs, the mass accretion rate drops monotonically in the dwarf
novae outbursts, and so does the fundamental frequency of the
oscillator, allowing an apparent frequency doubling as the higher
harmonics become resonantly excited.

The second high-frequency QPO in neutron stars does not have a
counterpart in accreting white dwarfs. The HF QPOs in black holes
cannot be excited by the spin of the (non-existent) central star,
their frequencies are not variable and are in a definite 2:3 ratio,
in agreement with the relativistic model of internal accretion disc
resonance.

\section*{Acknowledgements}

We are grateful to Tomaso Belloni for information about black-hole
HFQPOs. We thank the anonymous referee for his helpful remarks and
criticism. JPL was supported in part by a grant from the CNES.
Research supported in part by KBN grant 2P03D01424.

\label{lastpage}

\end{document}